\documentclass{PoS}

\def\sD{D \!\!\!\!/}

\title{An Overview of the Anomalous Soft Photons in Hadron Production}

\ShortTitle{An Overview of the Anomalous Soft Photons in Hadron Production}

\author{Cheuk-Yin Wong\thanks{Research supported in part by the Office of
    Nuclear Physics, US Department of Energy.}\\ 
    Oak Ridge National  Laboratory, USA\\ 
    E-mail: \email{wongc@ornl.gov}}

\abstract{ 
Productions of soft photon with low transverse momenta in high-energy
hadron-hadron collisions and $e^+$-$e^-$ annihilations indicate that
they are consistently produced in excess of what are predicted by
electromagnetic bremsstrahlung when hadrons (mostly mesons) are
produced, but they agree with electromagnetic bremsstrahlung
predictions in the absence of hadron production.  These excess soft
photons are called anomalous soft photons.  The occurrence of
anomalous soft photons in association with hadron production reveals
the presence of additional QED soft photon sources in QCD hadron
production.  Many different models have been proposed to explain the
anomalous soft production.  We shall examine specifically a quantum
field theory of simultaneous meson and soft photon production in
QCD$\times$QED in which the meson production arises from the
oscillation of color charge densities of the quarks of the underlying
vacuum in a flux tube.  As a quark carries both a color charge and an
electric charge, the oscillation of the color charge densities will be
accompanied by the oscillation of electric charge densities, which
will in turn lead to the simultaneous production of anomalous soft
photons during the meson production process.
}

\FullConference{International Conference on the Structure and the
Interactions of the Photon including the 20th International Workshop on
Photon-Photon Collisions and the International Workshop on High Energy
Photon Linear Colliders\\
                 20 - 24 May 2013\\
                 Paris, France}

\begin{document}

\section{Introduction}

In many exclusive measurements in the production of hadrons (mostly
mesons) since the 1980's \cite{Chl84}-\cite{DEL10}, it was found that
the production of associated soft photons with transverse momenta of
many tens of MeV is consistently greater than what is predicted from
the Low Theorem \cite{Low58} of electromagnetic bremsstrahlung.  The
excess low-$k_T$ soft photons are called anomalous soft photons.  They
have been observed in high-energy $K^+ p$ \cite{Chl84,Bot91}, $\pi^+
p$ \cite{Bot91}, $\pi^- p$ \cite{Ban93,Bel97,Bel02a}, $pp$ collisions
\cite{Bel02}, and $e^+$-$e^-$ annihilations in $Z^0$ hadronic decay
\cite{DEL06}-\cite{DEL10}.  They are absence when hadrons are not
produced \cite{DEL06}.

Theoretical treatment of the anomalous soft photons associated with
hadron production is a difficult problem because it involves the hadron
production in QCD,  the soft photon production in QED, and the interplay between QCD and QED particle production processes.
The occurrence of anomalous soft photons in association with hadron
production reveals the presence of additional QED soft photon sources
in QCD hadron production.  As a definitive quantum theory of hadron
production is not yet at hand, the anomalous soft photon problem
provide an interesting window to examine non-perturbative aspects of
QCD and QED particle production processes.  

While many different theoretical models have been proposed to explain
the anomalous soft photons, a complete understanding is still lacking.
It is of interest to present an overview of the experimental data as
well as theoretical approaches in our efforts to comprehend the
phenomenon.

\section{The Low Theorem}

Because the Low Theorem \cite{Low58} plays an important role in
quantifying the excess of soft photon production, it is worthwhile to
review its theoretical foundation and its contents.

\hangindent=5.0cm \hangafter=0 
We consider the process $p_1+p_2 \to p_3 + p_4 + k$ where $p_i$
represents a hadron and its momentum, and $k$ represents a photon and
its momentum.  For the simplest case with neutral $p_2$ and $p_4$, and
charged $p_1$ and $p_3$ hadrons, the Feynman diagrams are shown in
Fig.\ 1.  The amplitude for the production of a photon with a
polarization $\epsilon$ is
\begin{figure} [h]
\vspace*{-3.1cm}\hspace*{0.4cm}
\includegraphics[scale=0.45]{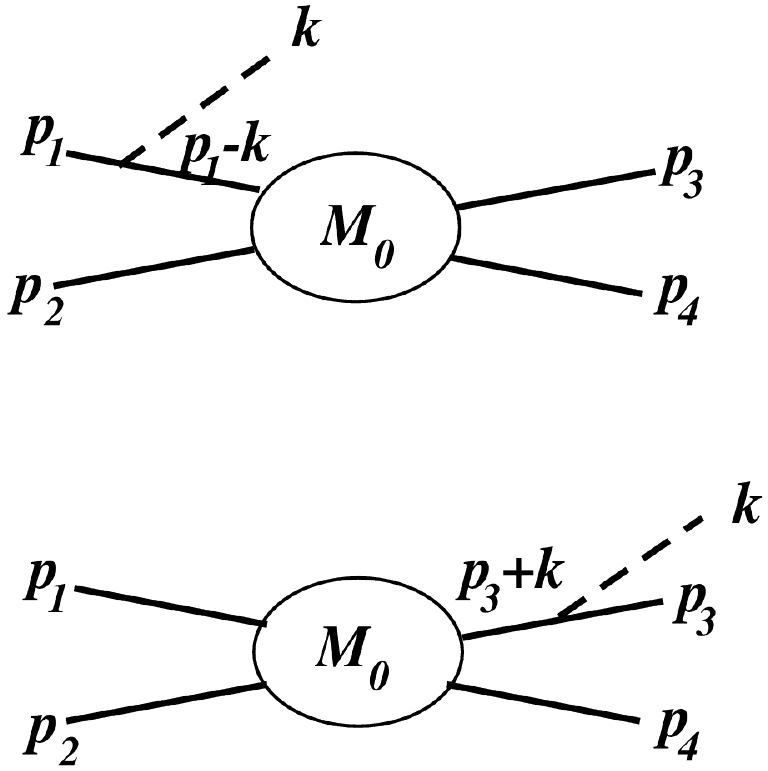}
\caption{ Feynman diagrams \hspace*{10.cm}~~~~~~~~~~\hfill\newline
for~$p_1$+$p_2$$\to$$ p_3$+$ p_4$+$ k$.\hspace*{5.0cm}~~~~~~~~~~~~~~~~~~~~~~~~~~~~~\hfill
  }
\end{figure}
\vspace*{-2.8cm}
\begin{eqnarray}
M(p_1p_2; p_3p_4 k)& =& M_0(p_1p_2; p_3p_4 )\left (\frac{e_1 p_1\cdot \epsilon}{(p_1-k)^2}
+\frac{e_3 p_3\cdot \epsilon}{(p_3+k)^2}\right ) ~~~~~~~\nonumber \\
 &  &\hspace*{-1.1cm}=M_0(p_1p_2; p_3p_4 )\left (
\sum_i^{\rm all~charged~particles}\frac{\eta_i e_i p_i\cdot \epsilon}{2p_i \cdot k}
\right )\!,
\label{2.1}
\end{eqnarray}

\noindent
where $e_i$ is the charge of $p_i$, and $\eta_i$ is +1 for an outgoing
hadron and -1 for an incoming hadron respectively.  In obtaining the
above equation, we have assumed
\begin{eqnarray}
M_0(p_1-k~~ p_2;~ p_3 ~p_4)\sim  M_0(p_1 ~p_2; ~p_3+k  ~~p_4)\sim M_0(p_1 ~p_2; ~p_3 ~p_4).
\end{eqnarray}
This is a reasonable assumption in high energy processes in which
$|{\bf p_1}|$ and $|{\bf p_3}|$ in the C.M.  frame are much greater
than the transverse momentum of the soft photon, $k_T$.  The amplitude
$M_0$ with the production of a soft photon is then approximately
independent of $k$ and can be adequately represented by $M_0(p_1p_2;
p_3p_4 )$, the Feynman amplitude for the production of only hadrons.
We can generalize the above Eq.\ (\ref{2.1}) to the process $p_1+p_2
\to p_3 + p_4 +...+ p_N+ k$ where $p_i$ is a hadron and $k$ is a soft
photon.  The Feynman amplitude is
\begin{eqnarray}
M(p_1p_2; p_3p_4... p_N k)& =& M_0(p_1p_2; p_3p_4...p_N )\left (
\sum_i^{\rm all~charged~particles}\frac{\eta_i e_i p_i\cdot \epsilon}{2p_i \cdot k}
\right ),
\end{eqnarray}
From the relation between Feynman amplitudes and cross sections, the
above equation gives \cite{Low58}
\begin{eqnarray}
\frac{  dN_\gamma}{ d^3 k}= \frac{\alpha}{2\pi k_0 } 
\int d^3p_1  d^3p_2d^3p_3 ... d^3p_N
\sum_{i,j=1}^{N}\eta_i \eta_j e_i e_j \frac{- (p_i \cdot p_j)} 
{(p_i \cdot k )(p_j \cdot k)}
\frac{dN_{\rm hadrons}}{d^3p_1  d^3p_2d^3p_3 ... d^3p_N}.
\label{22}
\end{eqnarray}
Thus, the spectrum of soft photons can be calculated from exclusive
measurements on the spectrum of the produced hadrons.

\section{
Experimental Measurements of Anomalous Soft Photon in Hadron production}

In which part of the photon spectrum are soft photons expected to be
important?  Upon choosing the beam direction as the longitudinal axis,
Eq.\ (\ref{22}) indicates that the contributions are greatest when the
transverse momentum of the photon, $k_T$, is small \cite{Gri67}.

\begin{table}[h]
\caption { The ratio of the soft photon yield associated with hadron
  production to the bremsstrahlung yield in high-energy hadron-hadron
  collisions and $e^+$-$e^-$ annihilations, compiled by
  V. Perepelitsa \cite{Per09}.  }
\vspace*{0.3cm}\hspace*{0.1cm}
\begin{tabular}{|l|l|c|c|}
\cline{1-4}
~~~~~~~~~~Experiment    &   Collision   &  Photon $k_T$    &     Photon/Brem
  \\
                     	&	Energy               &                           &    Ratio
 \\ \hline
     $K^+ p$, CERN,WA27,  BEBC (1984)          &~~70 GeV/c    &  $k_T <$ 60 MeV/c & 4.0 $\pm$0.8
 \\ \hline
     $K^+ p$, CERN,NA22,  EHS (1993)          & 250 GeV/c    &  $k_T <$ 40 MeV/c & 6.4 $\pm$1.6
 \\ \hline
     $\pi^+ p$, CERN,NA22,  EHS (1997)          & 250 GeV/c    &  $k_T <$ 40 MeV/c & 6.9 $\pm$1.3
 \\ \hline
     $\pi^- p$, CERN,WA83,OMEGA (1997)          & 280 GeV/c    &  $k_T <$ 10 MeV/c & 7.9 $\pm$1.4
 \\ \hline
      $\pi^+ p$, CERN,WA91,OMEGA (2002)          & 280 GeV/c    &  $k_T <$20 MeV/c & 5.3 $\pm$0.9
 \\ \hline
      $p p$, CERN,WA102,OMEGA (2002)          & 450 GeV/c    &  $k_T <$20 MeV/c & 4.1 $\pm$0.8 
 \\ \hline
      $e^+$$e^-$$ \to$hadrons, CERN,DELPHI   & $\sim$91 GeV(CM)    &  $k_T <$60 MeV/c &~ 4.0   
 \\ 
      with hadron production (2010)       &  &   & 
 \\ \hline
      $e^+$$e^-$$ \to$$\mu^+$$\mu^-$, CERN,DELPHI   & $\sim$91 GeV(CM)    &  $k_T <$60 MeV/c &~ 1.0   
 \\ 
      with no hadron production (2008)       &  &   & 
 \\ \hline
\end{tabular}
\end{table}

Many high-energy experiments were carried out to measure the spectrum
of photons with transverse momenta of the order of many tens of MeV.
The soft photon yields are then compared with what are expected from
electromagnetic bremsstrahlung of the hadrons as given by
Eq.\ (\ref{22}).  The results are summarized by Perepelitsa in Table I
\cite{Per09} and reviewed in the comprehensive report in \cite{DEL10}.
The experimental measurements indicate that low-$k_T$ soft photons are
produced in excess of what is expected from the electromagnetic
bremsstrahlung process.  In particular, in DELPHI measurements in
high-energy $e^+$-$e^-$ annihilations in $Z^0$ hadronic decay, the
ratio of the soft photon yield to the bremsstrahlung yield associated
with hadron production is about 4 \cite{DEL10}, whereas the ratio of
soft photon yield to the bremsstrahlung yield in the corresponding
$e^+ +e^- \to \mu^+ +\mu^-$ reaction is about 1 \cite{DEL06}.  This
indicates clearly that anomalous soft photons are present only when
hadrons are produced.

\section{Recent DELPHI Anomalous Soft Photon Data}
 
In addition to measuring the overall ratio of soft photon production,
the DELPHI Collaboration carried out measurements of soft
photon yields in different phase space in coincidence with various
hadron production variables.  They provide a wealth of information on
the characteristics of the produced hadrons associated with the
anomalous soft photons \cite{DEL06}-\cite{DEL10}.  The main
features of the observations from the DELPHI Collaboration can be summarized as
follows:
\begin{enumerate}
\item
Anomalous soft photons are produced in association with hadron
production at high energies.  They are absent when there is no hadron
production \cite{DEL08}.
\item
The anomalous soft photon yield is proportional to the hadron yield.
\item
The transverse momenta of the anomalous soft photons are in the region
of many tens of MeV/c.
\item
The anomalous soft photon yield increases approximately linearly with
the number of produced neutral or charged  particles, but,
the anomalous soft photon  yields increases  much faster with
increasing neutral particle multiplicity than with charged particle
multiplicity, as shown in Fig.\ 2.
\end{enumerate}
\begin{figure} [h]
\vspace*{-0.6cm}\hspace*{2.5cm}
\includegraphics[scale=0.40]{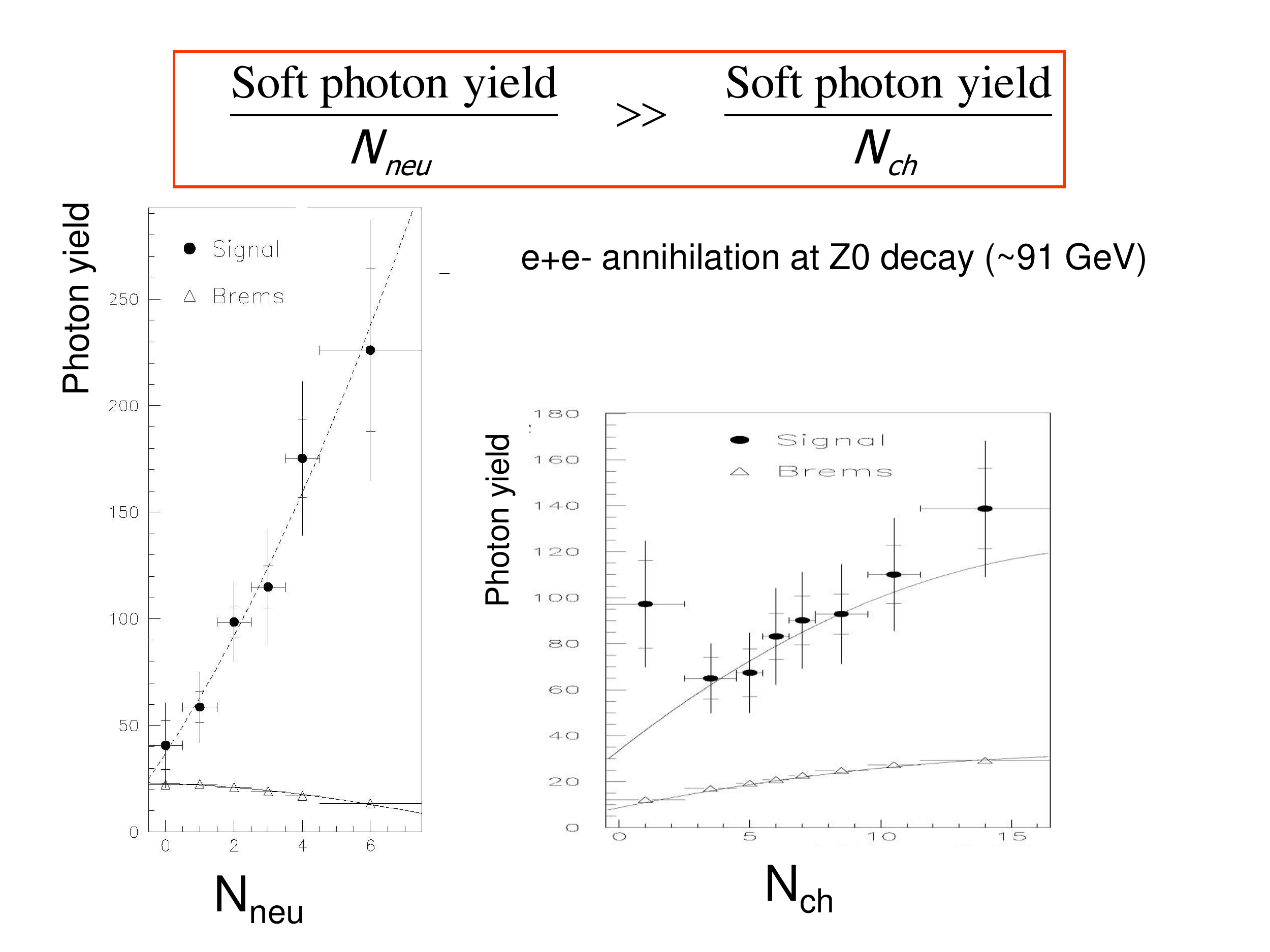}
\vspace*{-0.3cm}
\caption{ 
Number of soft photons $N_\gamma$ as a function of neutral hadron
multiplicity $N_{\rm neu}$ and charged hadron multiplicities 
$N_{\rm ch}$ from the DELPHI Collaboration \cite{DEL10}.  }
\end{figure}

\section{Models of Anomalous Soft Photon Production}

Many different model have been proposed to explain the anomalous soft
photon production in association with hadron production.  As the
transverse momenta of the anomalous photons are of order of many tens
of MeV, Van Hove and Lichard suggested that there is a source of these
low-energy photons in the form of a glob of cold quark-gluon system
of low temperature with $T \sim$10 to 30 MeV at the end of parton
virtuality evolution in hadron production \cite{Van89}.  In such a
cold quark-gluon plasma, soft photons may be produced by $q+\bar q \to
\gamma +g$ or $g+q \to \gamma + q$, and will acquire the
characteristic temperature of the cold quark-gluon plasma.  Kokoulina
$et~al.$ followed similar idea of a cold quark-gluon plasma as the
source of soft photon production \cite{Kok07}.  Collaborative evidence
of a cold quark-gluon plasma with $T\sim$ 10 to 30 MeV from other
sources however remain lacking.

Barshay proposed that pions propagate in pion condensate and they emit
soft photons during the propagation.  Rate of soft photon emission
depends on the square of pion multiplicity \cite{Bar89}.  The concept
of a pion condensation in high-energy $e^+$-$e^-$ annihilations in
$Z^0$ hadronic decay has however not been well established.  Shuryak
suggested that soft photons are produced by pions reflecting from a
boundary under random collisions. Hard reflections lead to no effect,
but soft pion collisions on wall leads to large enhancement in soft
photon yield \cite{Shu89}.  Czyz and Florkowski proposed that soft
photons are produced by classical bremsstrahlung, with parton
trajectories following string breaking in a string fragmentation.
They suggested that photon emissions along the flux tube agree with
the Low limit whereas photon emissions perpendicular to the flux tube
are enhanced over the Low limit \cite{Czy94}.  Nachtmann $et~ al.$
suggested that soft photons produced by synchrotron radiation from
quarks in the stochastic QCD vacuum \cite{Nac94}.  Hatta and Ueda
suggested that soft photons are produced in ADS/CFT supersymmetric
Yang-Mills theory \cite{Hat10}.  Simonov suggested that soft photons
arose from closed quark-antiquark loop \cite{Sim08}.
 
Wong proposed that soft photons arises in conjunction with the
production of hadron in the quantum field theory of particle
production in QCD2 and QED2 \cite{Won10}.  Such a mechanism will be
discussed in more detail in subsequent sections.  Following similar
lines of quantum field theory in two-dimensions, Kharzeev and Loshaj
suggested that soft photons arises from the induced current in Dirac
sea \cite{Kha13}.

\section{Quantum Field Theory of Simultaneous Meson and Soft Photon  Production}

While the various proposed models may explain some features of the
soft photon production process, the fourth feature listed in Section 4
from recent DELPHI observations cannot be explained by almost all
existing models \cite{Per09,DEL10}.  As the DELPHI data reveals
differential properties associated with the charged and neutral hadron
multiplicities and these multiplicities depend on the isospin
properties of the produced hadrons, we seek a quantum field theory
description for the simultaneous production of hadrons and anomalous
soft photons with explicit flavor degrees of freedom in $q$-$\bar q$
string-fragmentation that may explain this peculiar feature of the
phenomenon \cite{Won10}.

As described by Casher, Kogut, and Suskind \cite{Cas74}, the meson
production in the quantum field theory arises from the oscillation of
color charge densities of the quark vacuum in the flux tube when a
quark and an antiquark (or a diquark) pull away from each other at
high energies.  The meson rapidity distribution of these produced
exhibits a rapidity plateau, whose width increases with energy as
$\ln(\sqrt{s})$.  These color charge density oscillations obey
Klein-Gordon equations characterized by different masses of the mesons
\cite{Cas74}-\cite{Won09a}.  Because a quark carries both a color
charge and an electric charge, the underlying dynamical motion of the
quarks in the vacuum that generate color charge density oscillations
will also generate electric charge density oscillations in the flux
tube.  These color charge density oscillations will lead to the
production of photons that are additional to those from the
electromagnetic bremsstrahlung process.  Thus the oscillation of the
quark densities in the vacuum will lead to the simultaneous color and
electric charge density oscillations and will subsequently lead to
simultaneous and proportional production of mesons and anomalous
photons in the flux tube, in agreement with the first two features of
the anomalous soft photon phenomenon listed in Section 4.

It is of interest to examine whether the model also leads to results
that will be consistent with the other features of the anomalous soft
photon production phenomenon listed in Section 4.  For such a purpose, we study
quarks interacting with both QCD and QED interactions in the U$(3)$
group which comprises of the QCD color SU(3) and the QED
electromagnetic U(1) subgroups with different coupling constants.  We
introduce the generator $t^0$ for the U$(1)$ subgroup,
\begin{eqnarray}
t^0=\frac{1}{\sqrt{6}}
\left ( \begin{matrix}  
                {1 & 0 & 0 \cr
                0 & 1 & 0 \cr
                0 & 0 & 1 }
        \end{matrix}   \right ),
\end{eqnarray}
which adds on to the eight generators of the SU$(3)$ subgroup,
$\{t^1,...t^8\}$, to form the nine generators of the U(3) group.  They
satisfy $ {\rm tr}\{ t^\alpha t^ \beta \} = \delta^{\alpha\beta}/2
{~~\rm for~~} \alpha,\beta=0,1,..,8$.

We examine the QCD4$\times$QED4 system with two flavors in
four-dimensional space-time $x^\mu$.  The dynamical variables are the
quark fields, $\psi_f^a$, and the U(3) gauge fields,
$A_\nu$=$A_\nu^\alpha t^\alpha $, where $a$ is the color index, $f$
the flavor index, and $\alpha$ the U(3) generator index with coupling
constants $g_f^\alpha$,
\begin{eqnarray}
\label{qcdcc} 
g_u^{\{1,..,8\}}=g_d^{\{1,..,8\}}=g_{\rm QCD4}, {\rm~~for~~QCD},
\end{eqnarray}
\begin{eqnarray} 
\label{qedcc}
g_u^0=-e_u=-Q_u e_{\rm QED4}, ~~~g_d^0=-e_d=-Q_d e_{\rm QED4}, ~~~Q_u=2/3,~~~Q_d=-1/3
{\rm~~for~~QED}.
\end{eqnarray}
The transverse confinement of the flux tube can be represented by
quarks moving in a transverse scalar field $m({\bf r})$ where $m({\bf
  r})=S({\bf r})$+(current quark mass $m_q$) and $S({\bf r})$ is the
confining scalar interaction arising from nonperturbative QCD.  The
equation of motion of the quark field $\psi$ is
\begin{eqnarray}
\label{quark1}
\left \{ i \sD - m({\bf r}) \right \} \psi =0,
\end{eqnarray}
where 
\begin{eqnarray}
i\sD = \gamma^\mu \Pi_\mu =  \gamma^\mu ( p_\mu + g A_\mu).
\end{eqnarray}
The equation of motion for the gauge field $A_\mu$ is
\begin{eqnarray}
\label{Max4}
D_\mu F^{\mu \nu} = \partial_\mu F^{\mu \nu} -i g
[A_\mu, F^{\mu\nu}]
= g j^{\nu},  
\end{eqnarray}
where
\begin{eqnarray}
\label{F2}
F_{\mu \nu} = \partial_\mu A_\nu - \partial_\nu A_\mu   
-i g [A_\mu, A_\nu],
\end{eqnarray}
\begin{eqnarray}
\label{F3}
F_{\mu \nu}=F_{\mu \nu}^\alpha t^\alpha, 
~~~ j^\nu= j_{f}^{\nu \, \alpha} t^\alpha, 
\end{eqnarray} 
\begin{eqnarray}
\label{F4}
j^{\nu \,\alpha} =2 ~{\rm tr}~ {\bar \psi}_f \gamma^\nu t^\alpha \psi_f.
\end{eqnarray} 
Because of the commutative properties of the $t^0$ generator, the
commutator terms in Eqs.\ (\ref{Max4}) and (\ref{F2}) give zero
contributions for QED.

In the problem of particle production at high energies, the momentum
scales for longitudinal dynamical motion of the leading $q$ and $\bar
q$ as well as those of quarks in the underlying vacuum are much
greater than the momentum scales for their transverse motion. Hence,
inside the flux tube, $A^1$ and $A^2$ can be approximately neglected
and $A^0$ and $A^3$ can be considered as a function of $(x^0,x^3)$
only.  Under the longitudinal dominance and transverse confinement,
the QCD4$\times$QED4 system can be approximate compactified as a
QCD2$\times$QED2 system with a quark transverse mass $m_T$, and the
flux tube can be idealized as a $q$-$\bar q$ string \cite{Won09}.  The
coupling constant in two-dimensional space-time $g_{\rm 2D}$ and
four-dimensional space-time $g_{\rm 4D}$ are approximately related by
\cite{Won09}
\begin{eqnarray}
\label{est}
g_{\rm 2D}^2 \sim  \frac{g_{\rm 4D}^2}{\pi R_T^2}.
\label{coup}
\end{eqnarray}
A formal formulation of the compactification in a flux tube can be
carried out using the action integral, with a more rigorous definition
of the coupling constant normalization \cite{Kos12}.

\section{Bosonization of QCD2$\times$QED2 for quarks with two flavors}

Subsequent to the compactification, we can work with the effective
theory of QCD2$\times$QED2 in two-dimensional space-time.  For brevity
of notation, the two-dimensional nature of various quantities will be
understood in what follows except specified otherwise. The Lagrangian
density for QCD2$\times$QED2 that corresponds to the two-dimensional
version of Eqs.\ (\ref{quark1})-(\ref{F4}) is
\begin{eqnarray}
\label{lag}
{\cal L}= {\bar \psi} [ \gamma^\mu(i\partial_\mu +g A_{\mu}) - m_T ]
\psi - \frac{1}{4} F_{\mu \nu} F^{\mu \nu}. 
\end{eqnarray}

To search for bound states arising from the density oscillations of
the color and electric quark charges in QCD2$\times$QED2, one can use
the method of bosonization in which bi-linear products of fermion
field variables can be represented by functions of boson field
variables.  The bosons are bound and nearly free, with residual
sine-Gordon interactions that depend on the quark mass
\cite{Col75,Col76},\cite{Man75}-\cite{Gro96}.  The bosonization
program consists of introducing boson fields to describe an element
$u$ of the U(3) group and showing subsequently that these boson fields
lead to stable bosons with finite or zero masses.

For the QED2 bosonization in electromagnetic U(1), an element of the
U(1) subgroup can be represented by the boson field $\phi^0$ as $u =
\exp\{ i 2 \sqrt{\pi} \phi^0 t^0 \} $.  Such a bosonization poses no
problem as it is an Abelian bosonization.  It will lead to a stable
boson as in Schwinger's QED2 \cite{Sch62}.

For the QCD2 bosonization in color SU(3), we need to introduce boson
fields to describe an element $u$ of SU(3) in terms of the eight
$t^\alpha$ generators, as represented in general by $u = \exp\{ i 2
\sqrt{\pi} \phi \sum_{\alpha=1}^8 n^\alpha t^\alpha\}$ with an
amplitude $\phi$ and the unit vector ${\bf n}$=$\{n^1,n^2,..,n^8\}$.
The bosonization of the fermion current depends on $u^{-1}$ and
$\delta u/\delta x^\pm $ \cite{Wit84}.  A variation of $\delta
u/\delta x^\pm $ in any of the seven orientation angles of ${\bf n}$
will lead to $\delta u/\delta x^\pm $ quantities along other
$t^\alpha$ directions.  These variations of $\delta u/\delta x^\pm $
will not in general commute with $u$ or $u^{-1}$.  They will lead to
$j_\pm$ currents that are complicated non-linear functions of the
eight generator degrees of freedom and will make it difficult to look
for stable boson states with these currents.  On the other hand, a
variation of the amplitude $\phi$ in $u$, with the orientation of
${\bf n}$ held fixed, will lead to $\delta u/\delta x^\pm $ that
commutes with $u$ and $u^{-1}$ in the bosonization formula, as in the
Abelian case.  It will lead to simple currents and stable QCD2 bosons
with well defined masses.  We are therefore well advised to search for
stable bosons by varying only the amplitude of $\phi$, keeping the
orientation of ${\bf n}$ fixed.  As a unit vector ${\bf n}$ in any
orientation can be rotated to the first axis along the $t^1$ direction
by an orthogonal transformation in the $\alpha$-space, we can consider
the unit vector ${\bf n}$ to lie along the $t^1$ direction for the
bosonization of the color SU(3) subgroup.  We are therefore justified
to bosonize an element $u$ of the U(3) group as
\begin{eqnarray}
\label{uua}
u = \exp\{ i 2 \sqrt{\pi} \sum_{\alpha=0}^1 \phi^\alpha t^\alpha\}.
\end{eqnarray}
With such a bosonization, the kinetic energy, the interaction energy,
and the mass term lead to the Hamiltonian density \cite{Won10}
\begin{eqnarray}
\label{hh}
{\cal H}&=& \frac {1}{2}N_\mu \sum_{\alpha=0}^1 \biggl \{ \sum_{f=u,d}
 \left [ \frac{1}{2} (\Pi_f^\alpha)^2 + \frac{1}{2}
 (\partial_1\phi_f^\alpha )^2 \right ] + \frac{1}{ 2\pi} (\sum_{f=u,d}
 g_f^\alpha \phi_f^\alpha)^2 \biggr \} \nonumber\\ 
& & - \frac{e^\gamma m_T \mu}{2\pi}2 N_\mu \sum_{f=u,d} 
 \cos(2\sqrt{\pi/6}\phi_f^0)  \cos(2\sqrt{\pi/4}\phi_f^1),
\end{eqnarray}
where for $N_\mu$ is the normal order operator for the mass scale
$\mu$ \cite{Col76,Hal75}.

In the flavor sector, the up and down quarks combine to form the
isoscalar $(I,I_3)$=$(0,0)$ and the isovector $(I,I_3)$=(1,1), (1,0),
(1,-1) states.  The $(I,I_3)$=$(1,\pm1)$ QED2 states involve composite
constituents with like electric charges and are unlikely to be stable
in the electromagnetic sector.  We shall therefore focus our attention
only on the QED2 neutral isoscalar boson ($I$=0,$I_3$=0) and neutral
isovector boson ($I$=1,$I_3$=0) states.  In QCD with two flavors,
there is an approximate isospin symmetry, and the QCD quark-antiquark
meson states have isospin quantum numbers $I$ with nearly degenerate
$2I$+1 $I_3$-components.

We can construct the $\phi_I^\alpha$ fields for the isospin
$(I,I_3=0)$ states, for up and down quark fields moving in phase or
out of phase,
\begin{eqnarray}
\phi_I^\alpha = \frac{1}{\sqrt{2} } \left [ \phi_u^\alpha +(-1)^I
\phi_d^\alpha \right ]~~{\rm and}~~~
\Pi_I^\alpha = \frac{1}{\sqrt{2} }\left [ \Pi_u^\alpha +(-1)^I 
\Pi_d^\alpha \right ].
\end{eqnarray}
The Hamiltonian density for boson fields of different isospin quantum
numbers $I$ and $I_3=0$ is
\begin{eqnarray}
\label{ham0}
{\cal H}= \frac{1}{2}   N_\mu \biggl \{ 
   \sum_{\alpha=0}^1 
\sum_{I=0}^1 \biggl [\frac{1}{2}  (\Pi_I^\alpha)^2 
               + \frac{1}{2}  (\partial_1 \phi_I^\alpha)^2        
\biggr ] +  V (\{\phi_I^\alpha\}) 
                              \biggr \},
\end{eqnarray}
where $ V (\{\phi_I^\alpha\})= V_{\rm int}(\{\phi_I^\alpha\}) + V_{\rm
  m}(\{\phi_I^\alpha\})$ with the interaction energy
\begin{eqnarray}
 V_{\rm int} (\{\phi_I^\alpha\}) 
= 
\frac{1}{2}\left ( \sum_{I=0}^1 \frac{g_u^\alpha + (-1)^I g_d^\alpha}
                    {\sqrt{2}\pi} \phi_0^\alpha \right )^2,        
\end{eqnarray}
and the quark mass term 
\begin{eqnarray}
 V_{\rm m} (\{\phi_I^\alpha\}) 
=- \frac{e^\gamma m_T \mu}{2\pi} 2 \left [ \prod_{I=0}^1
\cos\left (\sqrt{2\pi}(\frac{\phi_I^0}{\sqrt{6}} +\frac{\phi_I^1}{\sqrt{4}})
\right )
+ \prod_{I=0}^1 \cos \left (\sqrt{2\pi}(\frac{\phi_I^0}{\sqrt{6}}
-\frac{\phi_I^1}{\sqrt{4}}) \right ) \right ].
\end{eqnarray}
The Hamiltonian density (\ref{ham0}) represents a QCD2 and QED2 system
of isoscalar and isovector boson fields $\phi_I^\alpha$ whose field
quanta acquire the mass $M_{I({\rm 2D})}^\alpha$, which can be
evaluated as the second derivatives of the potential at the potential
minimum located at $\phi_0^\alpha=0$ and $\phi_1^\alpha=0$.  We obtain
the mass square $(M_{I({\rm 2D})}^\alpha)^2$ of stable boson quanta
\begin{eqnarray}
\label{pot}
(M_{I({\rm 2D})} ^\alpha)^2
=\biggl [ \frac{\partial^2}{\partial (\phi_{I}^\alpha)^2} 
V(\{\phi_I^\alpha\})\biggr ]_{\phi_0^\alpha,\phi_1^\alpha=0} 
=\left ( \frac{g_u^\alpha+(-1)^I g_d^\alpha}{\sqrt{2\pi}} \right )^2
+ \frac{2}{3-\alpha} e^\gamma m_T \mu,
\label{mass}
\end{eqnarray}
where $\alpha=0$ for QED2 and $\alpha=1$ for QCD2.  As the boson field
$\phi_I^\alpha$ is related to the $\alpha$-gauge field, the quanta of
$\phi_I^\alpha$ are also the quanta of the $\alpha$-gauge fields.  The
QCD2 bosons and QED2 bosons can be appropriately called QCD2 mesons
and QED2 photons respectively.  Because $(M_{I({\rm 2D})}^\alpha)^2\ge
0$, the QCD2 mesons and QED2 photons are stable bosons.  The
self-consistency of gauge field oscillations and the induced quark
density oscillations lead to an equation of motion for the gauge field
oscillations in the form of Klein-Gordon equations characterized by
the corresponding masses.
 
If $(m_T \mu) $=0 (which corresponds to the massless quark limit), the
QCD2$\times$QED2 bosons are free.  With a finite value of $(m_T \mu)$,
they interact with a sine-Gordon residual interaction whose strength
depends on $(m_T \mu)$.  

\section{ QCD2 Meson and QED2 Photon Masses for Quarks with Two Flavors}

Equation (\ref{mass}) gives the QCD2 meson and QED2 photon masses for
quarks with two flavors.  As given by Eq.\ (\ref{coup}), the QCD2 and QED2 
coupling constants for quarks in a flux tube are
\begin{eqnarray}
\label{R1}
g_{\rm QCD2}^2 \sim 
\frac{g_{\rm QCD4}^2} {\pi R_T^2} \sim
\frac{4\alpha_s}{R_T^2},~~~{\rm and}~~~e_{\rm QED2}^2 \sim 
\frac{e_{\rm QED4}^2} {\pi R_T^2}
\sim \frac{4\alpha}{R_T^2},
\end{eqnarray}
where $\alpha_s=g_{\rm QCD4}^2/4\pi$ is the strong interaction
coupling constant and $\alpha=e_{\rm QED4}^2/4\pi=1/137$ is the fine
structure constant.  The magnitude of the flux tube radius $R_T$ is
revealed by the root-mean-squared transverse momentum of produced
hadrons (mostly pions) as
\begin{eqnarray}
\label{R2}
R_T \sim \frac{1}{\sqrt{\langle p_T^2\rangle_\pi } },
\end{eqnarray}
which empirically is slightly energy-dependent \cite{Won09a}.  We
shall focus our attention in high energy $e^+$-$e^-$ annihilations in
the hadronic decay of $Z^0$ at $\sqrt{s}\sim $91 GeV for which
$\sqrt{\langle p_T^2\rangle_\pi }=0.56$ GeV \cite{Bar97}, and thus
$R_T \sim 0.35$ fm.  We can infer the value of the nonperturbative
strong coupling constant $g_{\rm QCD2}$ from the string tension
coefficient $\kappa \sim 1$ GeV/fm that is related to $g_{\rm QCD2}$
by $g_{\rm QCD2}^2=2\kappa$ \cite{Won08,Won09,Won09a}.  We have
\begin{eqnarray}
g_{\rm QCD2} =   0.632 {\rm ~GeV}~~~{\rm and}~~~e_{\rm QED2}  \sim  0.096 {\rm ~GeV}.
\end{eqnarray}

With these coupling constants, ($g_u^1$=$g_d^1$=$g_{\rm QCD2}$,
$g_u^0$=$-Q_u e_{\rm QED2}$, and $g_d^0$=$-Q_d e_{\rm QED2})$, the
QCD2 and QED2 boson masses in the massless quark limit of $(m_T\mu)$=0
are shown in Table I.  One observes that QCD2 for quarks with two
flavors gives a massless pion in the massless quark limit, in
agreement with the concept of the pion being a Goldstone boson in the
standard QCD theory.  The massless isovector QCD2 meson lies lower
than the isoscalar QCD2 meson of 504 MeV, whereas the ordering is
opposite for the QED2 photons, with an isoscalar QED2 photon at 12.8
MeV and an isovector QED2 photon at 38.4 MeV.  These QED2 photons lie
in the region of observed anomalous soft photons.
\vspace*{-0.3cm}
\begin{table}[h]
  \caption { QED2 and QCD2 boson masses obtained with $R_T$=0.35 fm
and  $g_{\rm QCD2}^2$=$2b$=0.4 GeV$^2$.  }
\vspace*{0.2cm} 
\hspace*{-0.0cm}
\begin{tabular}{|c|c|c|c|}
\cline{3-4} \multicolumn{2}{c|}{} & QCD2 & QED2 
       \\  
\cline{1-4}\multicolumn{2} {|c|} {\hspace*{-0.8cm}Coupling Constant} 
& $g_{\rm QCD2}$=632.5 MeV & $e_{\rm QED2}$=96 MeV         \\ \cline{1-4} 
$(m_T\mu)$=0  & isoscalar boson mass  $M_{0({\rm 2D})}$ 
&  504.6 MeV &  12.8 MeV 
 \\  
        & isovector boson mass $M_{1({\rm 2D})}$
&  0         &  38.4 MeV 
 \\ \cline{1-4}
$m_T$=400 MeV & isoscalar boson mass $M_{0({\rm 2D})}$
&  734.6 MeV &  
        \\  
$\mu$=$m_T$~~~~~   & isovector boson mass $M_{1({\rm 2D})}$   
&  533.8 MeV &  
           \\ \cline{1-4}  
 $m_T$= 400 MeV  & isoscalar boson mass $M_{0({\rm 2D})}$
&            &  $O$(25.3 MeV) 
 \\  
$\mu$=$m_q$=$O$(1 MeV) & isovector boson mass $M_{1({\rm 2D})}$  
&                 &  $O$(44.1 MeV) 
  \\ \cline{1-4}  
\end{tabular}
\end{table}

Equation (\ref{pot}) indicates that the boson masses depend on four
mass scales: $g_{\rm QCD2}$, $e_{\rm QED2}$, $m_T$, and $\mu$.
Because a pion is a quark-antiquark composite, we can estimate the
quark transverse mass $m_T$ from the pion transverse momentum, $m_T
\sim \sqrt{\langle p_T^2\rangle_\pi /2}$ which gives $m_T \sim 0.4$
for $Z^0$ hadronic decay.  The mass scale $\mu$ arises from the
bosonization of the scalar density ${\bar \psi} \psi$ which diverges
in perturbation theory.  It has to be renormalized such that $\langle
{\bar \psi} \psi \rangle$=0 in a free theory.  It will need to be
renormal-ordered again in an interacting theory \cite{Col76}.  The
scalar density and the corresponding mass scale therefore depends on
the interaction.  For the strong interaction of QCD, confinement and
chiral symmetry breaking dominate and lead to a transverse mass $m_T$
that is much greater than the current quark mass.  It is reasonable to
take the mass scale $\mu$ to be the same as the quark transverse mass
$m_T$ in QCD2.  In QED2 with a relatively weak interaction, the scalar
density ${\bar \psi} \psi$ that diverges in perturbation theory has to
be renormalized in a nearly-free field in which the quark energy is
just the current quark masses.  The mass scale $\mu$ for QED2 should
therefore be the QED current quark mass $m_q$ appropriate for a
nearly-free theory.  We shall take $\mu=1$ MeV for QED2.

The values of the QCD2 and QED2 boson masses obtained with these
coupling constants and mass scales $\mu$ are given in Table I.  They
give a QCD2 isovector meson mass of 0.534 GeV and a QCD2 isoscalar
meson mass of 0.735 GeV in the flux tube and an isoscalar photon of
order 25 MeV and an isovector photon of order 44 MeV.  They fall
within the same order of magnitude of the transverse masses of
produced mesons and anomalous soft photons.

We can infer from the quantum field theory description of particle
production in Ref.\ \cite{Cas74} that the QCD2 mesons and QED2 photons
will be produced simultaneously in the same process of $q$-$\bar q$
string fragmentation, when a quark pulls away from an interacting
antiquark at high energies.  The production of a QCD2 meson of a
certain isospin quantum number will be accompanied by the production
of a QED2 photon of the same isospin, with appropriate probability
ratios.

In the emergence of the boson from two-dimensional space-time to
four-dimensional space-time, we envisage an adiabatic transverse
expansion from the two-dimensional flux tube to the full configuration
space.  The adiabatic transverse expansion involves no change of the
particle energy and particle longitudinal momentum so that the boson
mass $M_{I({\rm 2D})}^\alpha$ in two-dimensional space-time turns into
the boson transverse mass $ M_{IT}^\alpha$ in four-dimensional
space-time.  For example, we can consider the production of the lowest-mass
isovector meson, which is a pion.  Theoretically, Table I gives a
produced pion of mass $M_{I=1({\rm 2D})}^\alpha$=0.534 GeV in the flux
tube, corresponding to a boson of transverse mass 0.534 GeV in
four-dimensional space-time.  Experimentally, a pion is produced with
an average $\sqrt{\langle p_T^2 \rangle}_\pi \sim 0.56$ GeV
\cite{Bar97} for the $Z^0$ hadronic decay and the experimental
(average) isovector meson (pion) transverse mass is $M_{1T}^h$= 0.579
GeV, which is close to the value of 0.534 GeV in Table I.  We consider
next the production of the lowest mass isoscalar meson, which is the $\eta$ meson
with a rest mass $M_\eta=0.547$ GeV.  Theoretically, Table I gives a
produced isoscalar particle of mass $M_{I=0({\rm 2D})}^\alpha$=0.735
GeV in the flux tube, corresponding to a boson of transverse mass
0.735 GeV in four-dimensional space-time.  Experimentally, the average
transverse momentum of $\eta$ has however not been measured.  As the
$\eta$ meson has approximately the same mass as a kaon whose average
transverse momentum has been measured, the average transverse momentum
of the $\eta$ meson should be of the order of the kaon average
transverse momentum of $\sqrt{\langle p_T^2 \rangle_K }$$\sim$ 0.616
GeV \cite{Aih88}.  Upon taking this estimate to be the isoscalar meson
average transverse momentum, the experimental (extrapolated) isoscalar
meson ($\eta$ meson) average transverse mass is $M_{0T}^h$$\sim$ 0.824
MeV which can be compared well with $M_{\eta({\rm QCD2)}}$=0.735 GeV
in Table I.  In the case of QED photons, the rest mass of the photon
is zero in four-dimensional space-time.  Hence, the QED2 photon mass
in two-dimensional space-time can be identified as the photon
transverse momentum in four-dimensional space-time.  They fall within
the $k_T$ regions of anomalous soft photons.

\section{Rates of Meson and Anomalous Soft Photon Production}

To obtain an estimate on the rate of particle production, we shall use
the Schwinger mechanism of particle production in a strong field
\cite{Sch51}-\cite{Won95}. The probability of particle production of a
composite particle of transverse mass $M_{IT}^\alpha$ in an electric field of strength 
$\kappa_{q\bar
  q}^\alpha$
depends on the
exponential factor of $\exp\{-\pi (M_{IT}^\alpha/2)^2/ \kappa_{q\bar
  q}^\alpha \}$, where the factor of 1/2 in $M_{IT}^\alpha/2$ is to
denote the production of a pair of particles each of which has a mass
$M_{IT}^\alpha/2$, and combining one particle of mass
$M_{IT}^\alpha/2$ with the neighboring particle of mass
$M_{IT}^\alpha/2$ leads subsequently to a composite stable boson of
mass $M_{IT}^\alpha/2+M_{IT}^\alpha/2$.  Furthermore, from dimensional
analysis, we can deduce that the rate of production per space-time
volume element $(dz\, dt)$ has the dimension $\kappa_{q \bar
  q}^\alpha$.  We therefore infer from the Schwinger mechanism  that the rate of the production of
the number of particle of type $\alpha$, isospin $I$, and mass
$M_{IT}^\alpha$ due to the presence of the QCD and QED fields between
a receding quark $q$ and an antiquark $\bar q$ is
\begin{eqnarray}
\label{rate}
\frac{dN_I^\alpha} {dz\, dt} = A \sum_{q\bar q} P_{q\bar q}
~\kappa_{q\bar q}^\alpha ~\exp \left \{ -\frac{\pi
(M_{IT}^\alpha/2)^2}{\kappa_{q\bar q}^\alpha} \right \},  ~~~\alpha=\gamma,h,
\label{Sch}
\end{eqnarray}
where $P_{q\bar q}$ is the probability for the
quark-antiquark source pair to be a $u\bar u$ or $d \bar d$ pair, and
$A$ is a dimensionless constant.  In an $e^+$-$e^-$ annihilation at
high energies, there is an equal probability for the quark-antiquark
leading source pair to be a $u \bar u$ or $d \bar d$ pair, and so $P_{u\bar
u}=P_{d\bar d}=1/2$.

For the production of QCD2 mesons, the color electric field strength
between the leading quark and antiquark is independent of the quark
flavor quantum number.  It is given by
\begin{eqnarray}
\label{kappa}
\kappa_{u\bar u}^h=\kappa_{d\bar d}^h={g_{\rm QCD2}^2}/{2}=\kappa.
\end{eqnarray}
For the production of QED2 photons, the electric field strength
between the leading source quark $q$ and antiquark $\bar q$ is given
in terms of the electric charges of the quark and the antiquark as
\begin{eqnarray}
\kappa_{q \bar q}^\gamma = |Q_q Q_{\bar q}| e_{\rm QED2}^2/2.
\end{eqnarray}
Then the Schwinger mechanism of Eq.\ (\ref{Sch})  gives  
\begin{eqnarray}
\frac{N_{I=0}^h}{N_{I=1}^h} \sim \frac{1}{4},~~~
\frac{N_{I=0}^\gamma} {N_{I=1}^\gamma} \sim \frac{ 11}{4},  ~~~
\label{94}
\end{eqnarray}
which reveals that as the mass of an isoscalar meson is greater than
the mass of an isovector meson, as given in Table I, so isoscalar
mesons are much less likely produced than isovector meson (in a
particular $I_3$ state).  On the other hand, the mass of an isoscalar
($I$=0,$I_3$=0) photon is less than the mass of an isovector
($I$=1,$I_3$=0) photon, so isoscalar photons are much more likely
produced than isovector photon.  We can also calculate from
Eq.\ (\ref{94}) the ratio of ratios,
\begin{eqnarray}
\frac{N_{I=0}^\gamma} {N_{I=0}^h} : \frac{N_{I=1}^\gamma} {N_{I=1}^h} 
\sim \frac{ 11}{4} : \frac{1}{4} = 11 : 1,
\label{95}
\end{eqnarray}
which states that the number of soft isoscalar photons associated with
the isoscalar meson production are more numerous than soft isovector
photons associated with the isovector meson production.  The production of 
an isoscalar $\eta$ meson will lead to 1.64 neutral particles
and 0.57 charged particles while the production of an isovector
pion will lead, on the average,  to 2/3  charged particle and 1/3 neutral particle
\cite{PDG08}.  Production of an isoscalar meson is associated more
with the production of neutral mesons whereas the production of an
isovector meson is associated more with the production of charged
particles.  As a consequence,  the above ratio in Eq.\ (\ref{95}) suggests that
the ratio of the number of soft photons  to  neutral meson multiplicity
is much greater than the ratio of the number of soft photons  to  charged meson multiplicity,
in qualitative agreement with the
fourth feature in Section 4 of the DELPHI observation.

\section{Conclusions and Discussions}

In high energy hadron-hadron collisions and $e^+$-$e^-$ annihilations,
soft photons are produced in excess of what is expected from
electromagnetic bremsstrahlung predictions, indicating the presence of
additional anomalous QED soft photon production sources in QCD hadron
production processes.

We review here various models of anomalous soft photon production.
The most intriguing observation from the DELPHI Collaboration is the
peculiar feature that the yield of anomalous soft photons increases at
a much greater rate with increasing neutral particle multiplicity than
with charged particle multiplicity.  Such a feature cannot be
explained by almost all existing models \cite{Per09,DEL10}.  We
examine in this paper a quantum field theory description of the
simultaneous production of hadron and anomalous soft photons in a flux
tube that may explain this peculiar feature of the phenomenon
\cite{Won10}.

In this description, a color flux tube is formed when a quark and an
antiquark (or a diquark) pull apart from each other at high energies.
The motion of the quarks in the underlying vacuum of the flux tube
generates color charge oscillations which lead to the production of
mesons.  As a quark carries both a color charge and an electric
charge, the color charge oscillations of the quarks in the vacuum are
accompanied by electric charge oscillations, which will in turn lead
to the simultaneous production of soft photons during the meson
production process.

To study these density oscillations, we start with quarks interacting
with both QCD and QED interactions in four-dimensional space-time in
the U(3) group which breaks into the color SU(3) and the QED U(1)
subgroups.  Specializing to particle production at high energies, we
find that the dominance of the longitudinal motion and transverse
confinement lead to the compactification from QED4$\times$QED4 in
four-dimensional space-time to QCD2$\times$QED2 in two-dimensional
space-time, with the formation of the flux tube.  In the flux tube, the
self-consistent coupling of quarks and gauge fields lead to color
charge and electric charge oscillations that give rise to stable QCD2
bosons and QED2 bosons.  The boson masses depend on the gauge field
coupling constants.  The presence of the flavor degrees of freedom
leads to isospin dependence of the boson masses, with the isovector
meson mass smaller than the isoscalar meson mass, but the mass
ordering is reversed for the isoscalar photon and the isovector
photon.

As QCD2 and QED2 bosons are stable in the flux tube environment, we
can infer from the quantum field theory description of particle
production in Ref.\ \cite{Cas74} that these QCD2 mesons and QED2
photons will be produced simultaneously in $q$-$\bar q$ string
fragmentation.  Under the condition of adiabaticity with no change of
the particle energy and longitudinal momentum after the produced
particle emerges from the production region, the boson mass in
two-dimensional space-time turns into the boson transverse mass in
four-dimensional space-time.

Because both color charge oscillations and electric charge
oscillations arise from the same density oscillations of the quarks in
the vacuum, both QCD2 meson and QED2 photon will be simultaneously
produced by the fragmentation of the $q$-$\bar q$ string.  As the QCD
isoscalar meson mass is greater than the isovector mass, the
production of isoscalar mesons is less likely than isovector mesons.
In contrast, as the isoscalar photon mass is lower than the isovector
photon mass, the production of isoscalar photons is more likely than
isovector photons.  Production of an isoscalar meson is associated
more with the production of neutral mesons whereas the production of
an isovector meson is associated more with the production of charged
particles.  As a consequence, the ratio $N^\gamma/N_{\rm neu}$ is much
greater than the ratio $N^\gamma/N_{\rm ch}$, as observed by the
DELPHI Collaboration \cite{Per09,DEL10}.

While the QCD2$\times$QED2 model appears to explain qualitatively the
main features of the experimental anomalous soft photon data, it is
desirable to carry out further experimental measurements to test the
model:

\begin{enumerate}

\item It will be of interest to measure the transverse momentum
  distribution of the soft photons with a finer $k_T$ resolution and
  greater precision for a given narrow range of photon rapidities.
  Qualitatively, we expect the production of photons with two
  different average transverse momenta, one with $\sqrt{\langle
    k_T^2\rangle }$ $\sim$ 25 MeV for the production of the isoscalar
  photon and one with $\sqrt{\langle k_T^2\rangle }$ $\sim$ 44 MeV for
  the production of the isovector photon.

\item It will be of interest to measure the transverse momentum
  distribution by selecting events with predominantly neutral
  particles and events with predominantly charged particles.  The
  former events will likely arise from the production of isoscalar
  mesons and QED2 isoscalar photons, with an average photon transverse
  momentum of $\sim$25 MeV, while the latter from the production of
  isovector mesons and the isovector photons, with an average photon
  transverse momentum of $\sim$50 MeV.

\item
The rapidity distribution of the produced photons should exhibit the
plateau structure, as expected of similar distributions in meson
production.  A measurement of the rapidity distribution will provide
useful additional information on the dynamics of soft photon
production.

\item Measurements of the properties of associated hadrons similar to
  those of the DELPHI Collaboration should be carried out with
  hadron-hadron collisions at high energies where anomalous soft
  photon production has been reported \cite{Chl84}-\cite{Bel02a}.

\end{enumerate}

\end{document}